\documentclass[iop,apj]{emulateapj}
\slugcomment{To appear in the Astrophysical Journal}

\begin{document}

\title{HII Regions, Embedded Protostars, and Starless Cores in Sharpless 2-157}
\author{Chian-Chou Chen\altaffilmark{1}, Jonathan P. Williams\altaffilmark{1}
and Jagadheep D. Pandian\altaffilmark{1,2}}
\altaffiltext{1}{Institute for Astronomy, University of Hawaii at Manoa, Honolulu, HI, 96822, USA; ccchen,jpw@ifa.hawaii.edu}
\altaffiltext{2}{Indian Institute of Space Science and Technology,
Valiamala, Trivandrum 695547, India; jagadheep@iist.ac.in}
\shorttitle{Protostars in Sh 2-157}
\shortauthors{Chen, Williams \& Pandian}

\begin{abstract}
We present arcsecond resolution 1.4mm observations of the 
high mass star forming region, Sharpless 2-157, that reveal the
cool dust associated with the first stages of star formation.
These data are compared with archival images at optical, infrared,
and radio wavelengths, and complemented with new arcsecond resolution
mid-infrared data.  We identify a dusty young HII region, numerous
infrared sources within the cluster envelope, and four starless
condensations.  Three of the cores lie in a line to the south
of the cluster peak, but the most massive one is right at the center
and associated with a jumble of bright radio and infrared sources.
This presents an interesting
juxtaposition of high and low mass star formation within the same
cluster which we compare with similar observations of other high mass
star forming regions and discuss in the context of cluster
formation theory.
\end{abstract}


\section{Introduction}
Decades of observations of the closest molecular clouds have revealed
much about the formation of isolated low mass stars and the stages of
individual core collapse, protostellar birth, disk accretion, and
eventual dissipation of the surrounding envelope
\citep{2009ApJS..181..321E,2007ARA&A..45..565M}.
Such regions, however, are not the typical birth environments of most
stars, as is clear from the statistics of stellar clusters
\citep{2003ARA&A..41...57L}
and as inferred from the number of high mass, ionizing stars in the
Galaxy and the expectation from the IMF that they be accompanied by
large numbers of lower mass stars \citep{1997ApJ...476..144M}.
There is also considerable cosmochemical evidence that our Sun
formed in a large cluster, in close proximity to a massive star
\citep{2011AREPS..39..351D, 2010ConPh..51..381W}.
A more complete understanding of our origins
requires studies of a broader range of stellar birth environments
including clusters and high mass star forming regions.

High mass stars, $M_\ast\gtrsim 8\,M_\odot$,
cannot form in exactly the same way that solar mass stars form because
spherically symmetric accretion at the thermal sound speed would
be turned back by radiation pressure \citep{1987ApJ...319..850W}.
Self-similar turbulence provides a way to extend the low mass
paradigm to higher masses by providing enhanced accretion at the
turbulent, rather than thermal, speed \citep{2003ApJ...585..850M}.
Infall onto the star may also slip through radiation bubbles via
Rayleigh-Taylor instabilities \citep{2009Sci...323..754K}.
\cite{1998MNRAS.298...93B} proposed the coalescence of low mass
stars as a means to bypass the radiation pressure problem but the
observed density of young stellar clusters does not appear high enough
for this to be viable. The main alternative to the picture of quasistatic
collapse of a massive core is the dynamic scenario of competitive accretion
of a cluster of stars in a dense clump \citep{2006MNRAS.370..488B}.
Here, stars at the center of the cluster potential well grow rapidly
to high masses, whereas stars in the outer parts grow more slowly.
Detailed, multi-wavelength observations are required to test
these different scenarios.

Massive stars evolve much faster than low mass stars, achieving hydrogen
fusion while still accreting material from their surroundings.
Together with their intrinsic rarity, this means that high mass
protostars are few and far between.  High resolution observations are
therefore required to distinguish them from their clustered surroundings.
Their high luminosities, however, make them readily apparent
and the ionized (HII) gas that they produce shortly after their birth
can be detected at radio wavelengths where Galactic extinction is negligible
and is an unambiguous signature of their nature as massive stars.

\begin{figure*}[!ht]
\figurenum{1}
\centering
\includegraphics[width=6.5in]{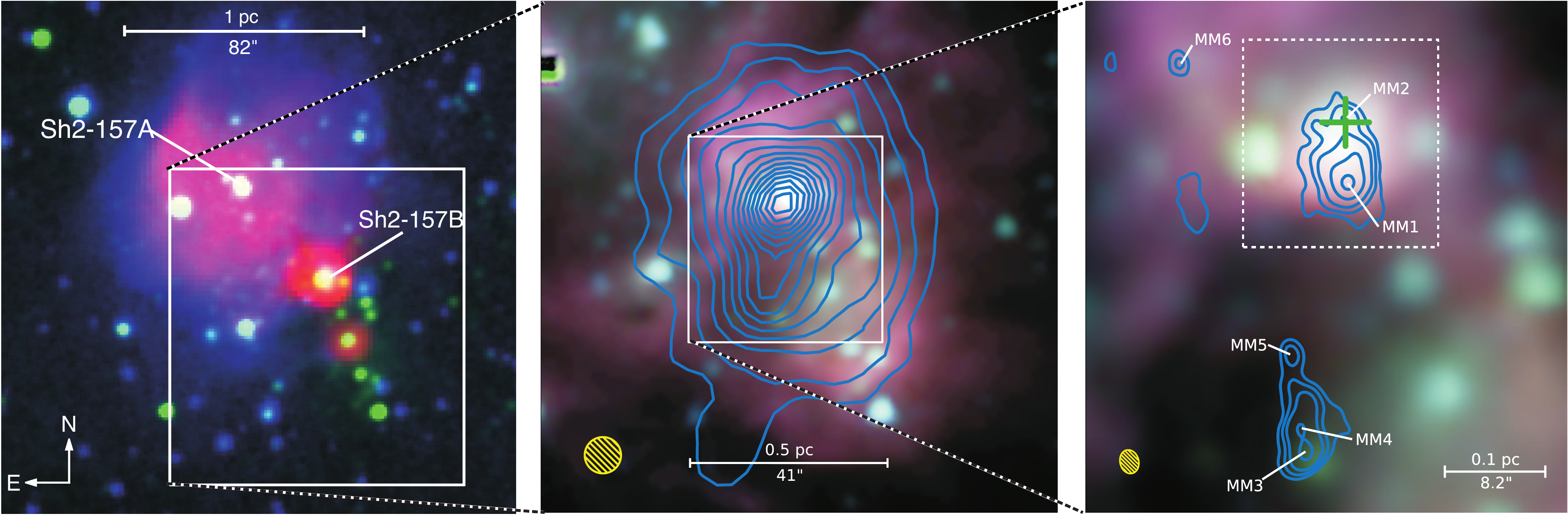}
\caption{The Sh2-157 HII massive star forming region over a range
of scales and wavelengths.  The leftmost panel presents the overview
of the entire region with red indicating VLA 3.6\,cm emission from
ionized gas, green indicating 2MASS K-band $2.2\,\mu$m emission mainly
from embedded protostars, and blue showing R-band optical nebulosity and
stars from the Digitized Sky Survey.
The box shows the region in the center panel which zooms onto
the ultra-compact HII region Sh2-157B and accompanying cool, dusty
envelope. The background three color image is Spitzer IRAC bands 1,2, 3
($3.6, 4.5, 5.8\,\mu$m respectively)
and the blue contour overlay shows SCUBA $450\,\mu$m emission.
The contour levels begin at, and increment in steps of, 1\,Jy\,beam$^{-1}$
and the $7\farcs 5$ SCUBA beamsize
is shown as the hashed circle in the lower left corner.
The box indicates the region in the rightmost panel which zooms
further onto Sh2-157B and the envelope ridge to the south.
The background image is the same Spitzer three color image and the
blue contour overlays are SMA 1.4\,mm emission with levels
at $[3,5,8,20,35]\times 1.6$\,mJy\,beam$^{-1}$.
The $1\farcs 8\times 1\farcs 5$ beamsize
is shown as the hashed ellipse in the lower left corner.
The green cross indicates the centroid of the VLA 3.6\,cm emission.}
\label{fig:zoom}
\end{figure*}

Here, we present a high resolution, multi-wavelength study of
the massive star forming region, Sh2-157.
First cataloged by \cite{1959ApJS....4..257S} on the basis of its optical
nebulosity, this well studied region has an estimated kinematic distance
of 2.5\,kpc \citep{1984ApJ...279..125F}.
The large HII region, Sh2-157A, is powered by a
late O star \citep{1972A&A....18..373C}
but radio observations by \cite{1973A&A....27..143I}
found a second, ultra-compact HII region, Sh2-157B, to the southwest
which is likely powered by an early B-star \citep{1994ApJS...91..659K},
and a third, fainter, HII region further to the
southwest \citep{1999ApJ...514..232K}.
A large scale radio-optical-near-infrared view of the region
is shown in the leftmost panel of Figure~\ref{fig:zoom}.
The optical nebulosity is centered in Sh2-157A and Sh2-157B lies
near the northern end of a string of infrared sources.

This region was imaged with the Spitzer Space Telescope as part of a
survey of ultra-compact HII regions (PI: Sean Carey).
Sh2-157B is a bright, extended source in these data.
It also lies at the peak of the SCUBA $450\,\mu$m
and $850\,\mu$m maps of \cite{2006A&A...453.1003T}.
These submillimeter data show the emission from cool dust in
a large cluster envelope.  The total integrated $450\,\mu$m flux is
151\,Jy which implies a total mass $\sim 700\,M_\odot$
(under standard assumptions that are spelled out in \S3.1).
We plot the archival IRAC and SCUBA data to show the embedded protostellar
population in the central panel of Fig~\ref{fig:zoom}.  
The infrared sources mostly lie along a line away from Sh2-157B
whereas the envelope extends in a more southerly direction.

The lack of infrared emission along the dense envelope ridge
suggests a mass reservoir for continued star formation.
In this paper, we present 1.4\,mm interferometric observations
of the cluster envelope to show its structure at comparable
resolution to the optical, infrared, and radio data.
We have also carried out ground-based mid-infrared observations
of Sh2-157B to obtain higher resolution images than the Spitzer data.
We find an elongated structure consisting of three low-mass starless cores
in the south and a tight mixture of protostellar evolutionary states in
the ultra-compact HII region at the center.  This provides new detail
into how massive stars and stellar clusters form and also
provides an interesting juxtaposition of high and low
mass star formation in adjacent regions.
Section 2 describes the data acquisition and reduction.
Section 3 describes the images and source characterization.
We discuss the implications of this work for massive star formation
in Section 4 and summarize in Section 5.

\section{Observations}
\label{sec:obs}
We used the Submillimeter Array\footnote[2]{The Submillimeter Array is
a joint project between the Smithsonian Astrophysical Observatory and the
Academia Sinica Institute of Astronomy and Astrophysics and is funded by
the Smithsonian Institution and the Academia Sinica.}
(SMA) to image the compact condensations within the
dusty envelope around Sh2-157B.
Observations were carried out in the extended configuration
(70--240 meter baselines) on 2008 August 18 and in the compact
configuration (20--70 meter baselines) on 2008 September 22.
A five pointing Nyquist sampled mosaic was made to provide uniform coverage
over the full region of SCUBA emission.

The shape of the bandpass was measured by observing 3C279 and 3C273.
The time dependent gains were determined by 5 minute observations of BLLac
between 20 minute observations on source.  The absolute flux scale
was determined from observations of Titan.
We used the data reduction package
MIR calibrate the visibilities and produced the images using
standard MIRIAD routines.
The 1.4\,mm continuum map was made from the line free channels and spanned
4\,GHz in bandwidth (lower plus upper sidebands).
The final map was made by inverting the uniformly weighted
compact and extended datasets together
and has an rms noise level of 1.3\,mJy\,beam$^{-1}$
and beamsize $1\farcs 8\times 1\farcs 5$.

The receivers were tuned to place the
H$_{2}$CO 3$_{1,2}$--2$_{1,1}$ line at 225.7 GHz in the upper sideband
and the DCN 3--2 line at 217.2 GHz in the lower sideband.
This was the same setup that proved useful in a similar study
of AFGL961 as diagnostics of infall, outflow, and cold
core chemistry \citep{2009ApJ...699.1300W}.
Both these lines were detected in Sh2-157B but we did not find clear evidence
for infall or outflow, and they are not used in the analysis described here.

\begin{figure*}[!ht]
\figurenum{2}
\centering
\includegraphics[width=6.0in]{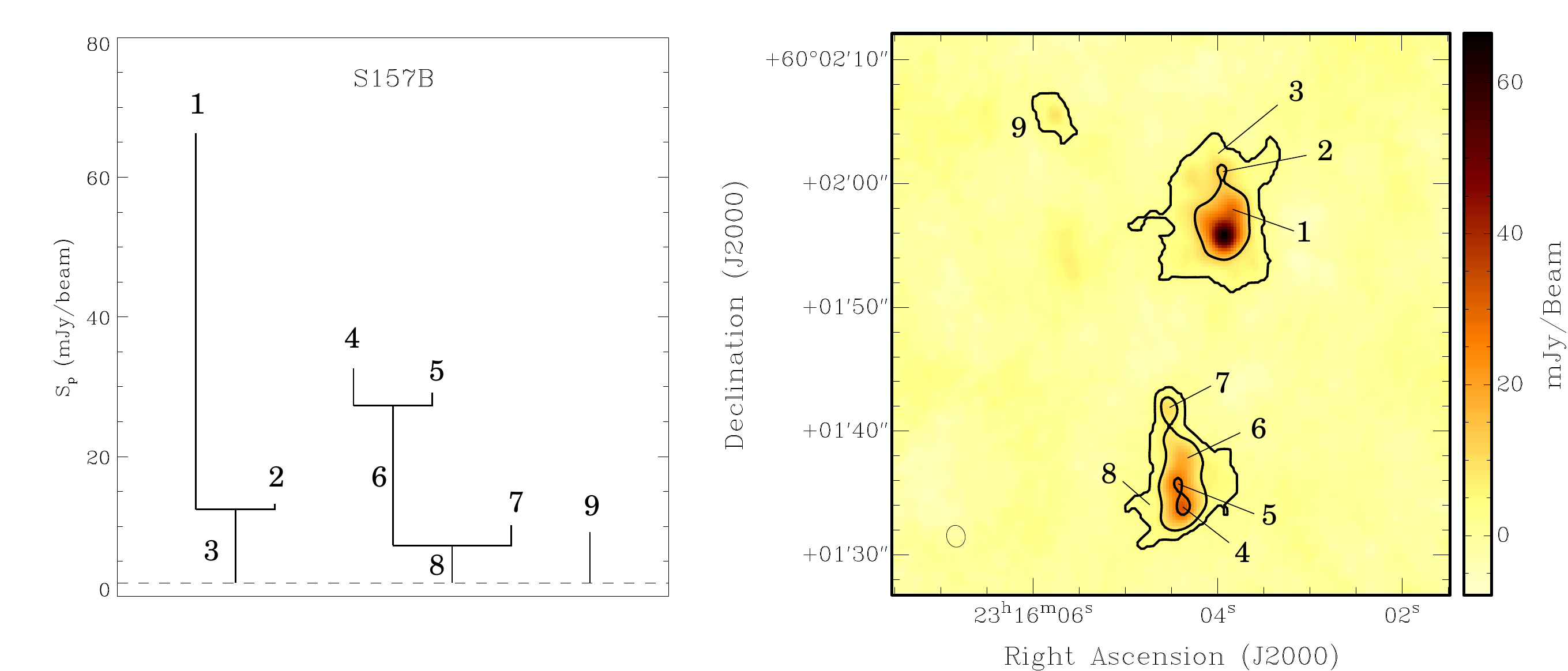}
\caption{
Dendogram structure decomposition of the SMA 1.4\,mm continuum
image of S157.  The left panel shows the dendogram tree and
the right panel labels each feature on the continuum map with contours
at break levels between different structures in the dendogram.
}
\label{fig:dendogram}
\end{figure*}

We used the mid-infrared camera, MIRSI \citep{2008PASP..120.1271K},
on the Infrared Telescope Facility\footnote[3]{Operated by the
University of Hawaii under Cooperative Agreement no. NNX-08AE38A
with the National Aeronautics and Space Administration,
Science Mission Directorate, Planetary Astronomy Program.} (IRTF)
to image Sh2-157B at higher resolution and at longer wavelengths
than the Spitzer IRAC data.
We observed in the N ($10.4\,\mu$m) and Q ($20.9\,\mu$m) filters,
on 2009 July 17 to 19 in dry, stable conditions.
The images were diffraction limited at
$0\farcs 9$ and $1\farcs 7$ resolution respectively.
We used a 10$''$ $\times$ 10$''$ dithering pattern
and also chopped off the chip with chopping 60$''$ north--south
and nodding 90$''$ east--west in order to obtain better sky
subtraction in this crowded area. Calibration was performed
via observations of $\beta$ Peg throughout the night.
The data reduction was performed with an IDL pipeline written by
staff astronomer Eric Volquardsen.  The plate scale and rotation
had been well determined from previous calibration observations
and the final images were registered to an
absolute astrometric grid by aligning with source
23160401+6002011 in the 2MASS Point Source Catalog that
is well detected in both MIRSI bands
(and lies at the northeast corner of the leftmost panel
of Figure~\ref{fig:zoom}.)

\section{Results}
\label{sec:results}
The SMA continuum emission is shown as contour overlays on the
Spitzer IRAC band 3 ($3.5\,\mu$m) image in the rightmost panel of
Figure~\ref{fig:zoom}.
The interferometer spatially filters out much of the envelope
and reveals the compact cores within.
As with all aspects of the interstellar medium, there is a hierarchy
of structures and, even with the spatial filtering, local peaks
lie within common regions of extended emission.
We first describe the decomposition of the map structure
and the resulting bounds to the masses of the compact features
and then discuss the relation of the two main groups of
sources in the context of their star forming nature.

\subsection{Structure decomposition of the millimeter emission}
There is an inherent ambiguity in the identification of irregularly
shaped objects that overlap in projection or share a common envelope.
Visual inspection of the SMA map shows a handful of local peaks
nested in two main groups of lower level emission.
Based on the nomenclature suggested by
\cite{2000prpl.conf...97W}, we consider these to be cores within clumps
with the connection to star formation being that cores may produce
individual stellar systems (single or small order close-multiple)
and that clumps are the potential progenitors of stellar groups.

To calculate the properties of the cores and clumps, and their
relation to each other, we first decompose the map into a tree network
or ``dendogram'' \citep{2008ApJ...679.1338R}.
This reveals nine features which are shown in relation to each in the
left panel of Figure~\ref{fig:dendogram}, and labeled in the image in
the right panel.  There are three distinct ``roots'', one of which
(feature 9) is a low level, isolated feature associated with diffuse
infrared emission.  The other two roots correspond to the two local peaks
in the SCUBA $450\,\mu$m map and labeled
G111.282--0.665SMM and G111.281--0.670SMM by \cite{2006A&A...453.1003T}.
The peaks at the end of these roots are ``leaves'' and are linked,
in one case, by an intermediate hierarchy or ''branch''.
Table~\ref{tab:dendogram} lists the nine features and their position in
the hierarchy, along with basic observational quantities; peak and
total intensities, and deconvolved area.  Of the six leaves,
only two are resolved.

The total intensity of a leaf in Table~\ref{tab:dendogram}
is the sum above the emission level at which the
feature is defined.  It therefore provides a minimum mass
to the core (this similarly applies to the size).
The mass may be considerably higher if we allow for lower
level emission from the branches and roots below.
The connection to the actual physical properties of the
core are less certain in this case but a well defined
algorithm, CLUMPFIND,  exists to allocate common envelope emission
to the local peaks within \citep{1994ApJ...428..693W}.
The dendogram analysis provides the emission levels at which the
different features separate (``nodes'' in the nomenclature of
\cite{2008ApJ...679.1338R}) -- these are the contours shown in
Figure~\ref{fig:dendogram} -- and we use these as inputs
into CLUMPFIND to extend the cores (or ``leaves'') to lower levels.
This effectively provides an upper limit to the total flux,
and therefore maximum mass, to each core.

\begin{deluxetable}{crcccc}
\tablewidth{0pt}
\tablenum{1}
\tablecaption{Dendogram Structure Analysis}
\label{tab:dendogram}
\tablehead{
\colhead{Feature} & \colhead{Type} & \colhead{$F_{\rm peak}$} & \colhead{$F_{\rm total}$} & \colhead{Area} \\
\colhead{} & \colhead{} & \colhead{(mJy/beam)} & \colhead{(mJy)} & \colhead{($\square^{''}$)}
}
\startdata
 1 & Leaf   & 66.3 & 177.4 & 20.0 \\
 2 & Leaf   & 13.3 &   2.9 & unresolved \\
 3 & Root   & 66.3 & 304.4 & 91.3 \\
 4 & Leaf   & 32.6 &  13.4 & unresolved \\
 5 & Leaf   & 29.2 &   6.4 & unresolved \\
 6 & Branch & 32.6 & 122.6 & 23.1 \\
 7 & Leaf   & 10.2 &   5.9 & unresolved \\
 8 & Root   & 32.6 & 180.2 & 63.4 \\
 9 & Leaf   &  9.2 &  13.0 &  9.3 \\[-2mm]
\enddata
\end{deluxetable}

\begin{deluxetable*}{lccccc}
\tablewidth{0pt}
\tablenum{2}
\tablecaption{Millimeter Cores}
\label{tab:mm_cores}
\tablehead{
\colhead{Core} & \colhead{Dendogram} & \colhead{$\alpha$(2000)} & \colhead{$\delta$(2000)} & \colhead{$F_{\rm total}$} & \colhead{$M_{\rm core}$} \\
\colhead{} & \colhead{Feature} & \colhead{(h:m:s)} & \colhead{(d:m:s)} & \colhead{(mJy)} & \colhead{($M_\odot$)}
}
\startdata
 Sh2-157B-MM1 & 1 & 23:16:03.93 & 60:01:56.0 & 177.4--239.2 & 12.5--17.0 \\
 Sh2-157B-MM2 & 2 & 23:16:03.96 & 60:02:01.0 &   2.9--25.9  &  0.2--1.8 \\
 Sh2-157B-MM3 & 4 & 23:16:04.36 & 60:01:34.0 &  13.4--63.9  &  0.9--4.5 \\
 Sh2-157B-MM4 & 5 & 23:16:04.43 & 60:01:35.8 &   6.4--46.8  &  0.5--3.3 \\
 Sh2-157B-MM5 & 7 & 23:16:04.53 & 60:01:42.0 &   5.9--14.4  &  0.4--1.0 \\
 Sh2-157B-MM6 & 9 & 23:16:05.76 & 60:02:05.5 &        13.0  &       0.9 \\[-2mm]
\enddata
\end{deluxetable*}

We calculate the millimeter emitting mass from the SMA 1.4\,mm
flux, $F_{\rm SMA}$, using the simple, optically thin, relation
$M=F_{\rm SMA}d^2/\kappa_\nu B_\nu(T)$, where $d=2.5$\,kpc
and $B_\nu(T)$ is the Planck function.
We assume a uniform temperature $T=20K$ and a dust opacity
$\kappa_\nu=0.1(\nu/1200\,{\rm GHz})$\,cm$^2$\,g$^{-1}$
\citep{1983QJRAS..24..267H} that implicitly includes a
gas-to-dust ratio of 100.
Although we do not know the dust temperature and opacity for
these particular objects, these are typical values for
protostellar cores as determined from detailed studies
of other regions \cite[e.g.,][]{1999ARA&A..37..311E}
and allow for direct comparisons.
The positions of the six cores,
labeled Sh2-157B-MM1 through Sh2-157B-MM6 and hereafter referred to
as MM1 though MM6,
are listed in Table~\ref{tab:mm_cores} along with the
range of millimeter fluxes and masses.
Because of the blending, the difference between the minimum and
maximum mass can be quite large but a clear distinction can nevertheless
be made.  That is, core MM1 has a mass in excess of $10\,M_\odot$
and the other five bracket a solar mass.
Their star forming nature is discussed for each clump
(dendogram roots 3 and 8) below.

\subsection{Protostars and cores in the northern clump}
\begin{figure}[tb]
\figurenum{3}
\centering
\includegraphics[height=5.0in]{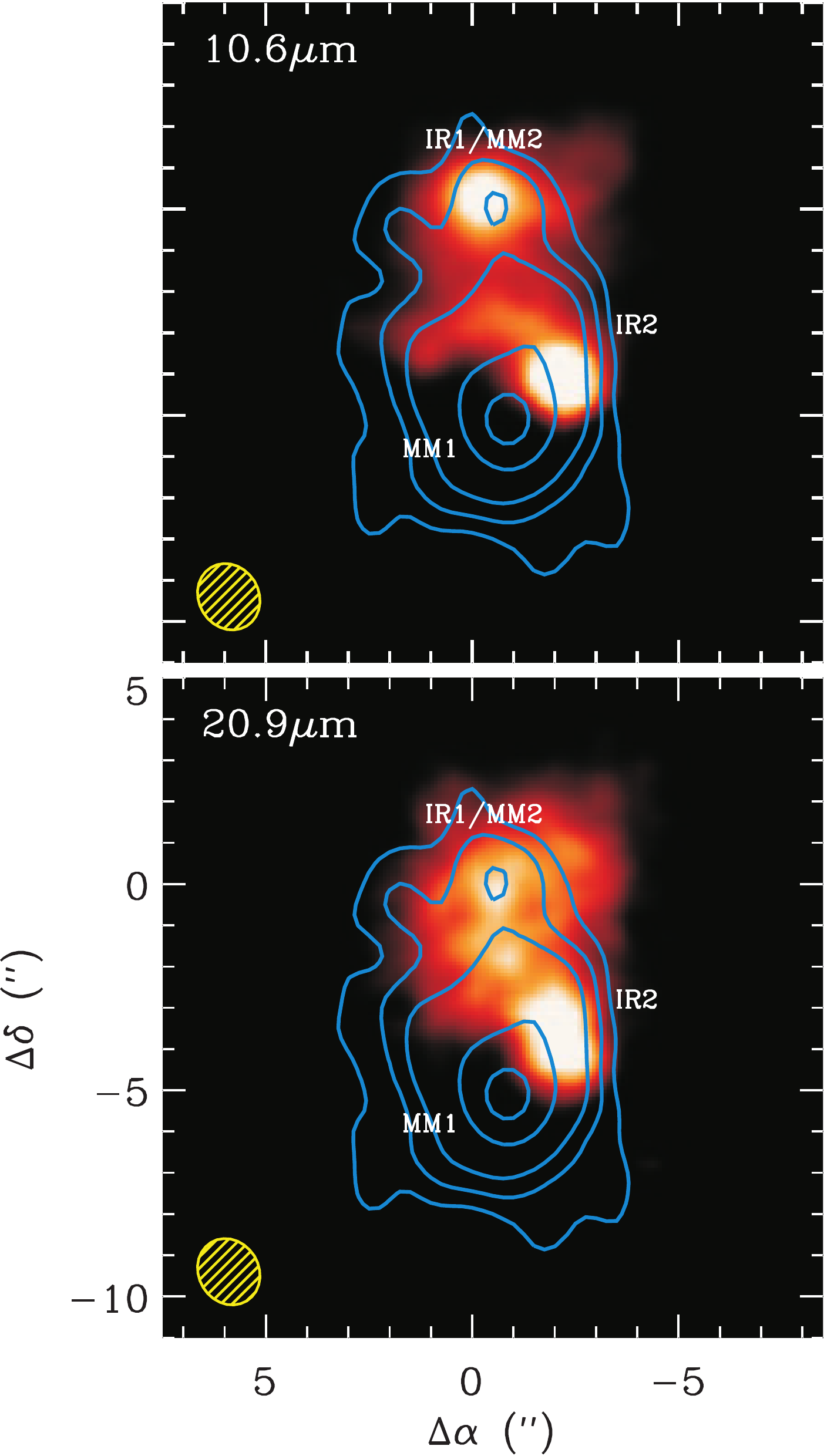}
\caption{Mid-infrared MIRSI images resolve Sh2-157B
region into two sources IR1 and IR2.  The origin of the maps is
$\alpha(2000)=23^{\rm h}16^{\rm m}04^{\rm s}.01,
\delta(2000)=60^\circ 02' 01\farcs 2$.
The top panel shows the N-band
($10.6\,\mu$m) image on a linear scale range from 3 to 10 mJy per
$0\farcs 27$ pixel.
The bottom panel shows the Q--band ($20.9\,\mu$m) image
on a linear scale range from 23 to 140 mJy per $0\farcs 27$ pixel.
Both panels include blue contour overlays of SMA 1.4\,mm emission
at $[3,5,8,20,35]\times 1.6$\,mJy\,beam$^{-1}$.
The blue hashed ellipse in the lower left corner of the bottom panel
is the $1\farcs 8\times 1\farcs 5$ SMA beamsize.}
\label{fig:mirsi}
\end{figure}

The northern clump,
labeled G111.282--0.665SMM by \cite{2006A&A...453.1003T},
is centered on the peak of the SCUBA maps.
It has a rising spectrum at infrared wavelengths and is saturated in
the Spitzer images in IRAC band 4 ($4.5\,\mu$m)
and MIPS bands 1 and 2 (24 and $70\,\mu$m).
Our MIRSI N and Q band (10 and $20\,\mu$m) images
therefore not only provide higher resolution than
the Spitzer data but also the first infrared photometry beyond $8\,\mu$m.
The MIRSI data are shown in Figure~\ref{fig:mirsi} 
with a contour overlay of the SMA emission.
There are two infrared sources,
Sh2-157B-IR1 and Sh2-157B-IR2 (hereafter IR1 and IR2),
at $10\,\mu$m and extended emission between them.
Both sources are also apparent at $20\,\mu$m but the
northernmost source, IR1, is diffuse rather than point-like at
this wavelength.
IR1 corresponds to the VLA source, i.e., the HII region Sh2-157B,
which is similarly extended at 2\,cm \citep{1994ApJS...91..659K},
and the SMA core MM2.
IR2 lies within the SMA contours but is significantly offset
from the emission peak, core MM1.
We conclude that there are three distinct sources in this region:
IR1/MM2 with infrared, millimeter, and radio emission,
the infrared source IR2,
and the massive core, MM1, that is detected only in the millimeter.

\begin{deluxetable*}{lccccccc}
\tablewidth{0pt}
\tablenum{3}
\tablecaption{Infrared Sources: Photometry}
\label{tab:ir_sources}
\tablehead{
\colhead{Source} & \colhead{$\alpha$(2000)} & \colhead{$\delta$(2000)} & \colhead{J} & \colhead{H} & \colhead{K} & \colhead{N} & \colhead{Q} \\
\colhead{} & \colhead{(h:m:s)} & \colhead{(d:m:s)} & \colhead{(Jy)} & \colhead{(Jy)} & \colhead{(Jy)} & \colhead{(Jy)} & \colhead{(Jy)}
}
\startdata
 Sh2-157B-IR1 & 23:16:03.96 & 60:02:01.4 & 0.015 & 0.027 & 0.034 & 1.0 & 33.5 \\
 Sh2-157B-IR2 & 23:16:03.71 & 60:01:57.0 & 0.002 & $<0.019\tablenotemark{a}$ & $<0.028$\tablenotemark{a} & 1.1 & 13.6 \\[-2mm]
\enddata
\tablenotetext{a}{IR2 is confused with the brighter IR1 at H and K band
and only upper limits to its flux can be determined.}
\end{deluxetable*}

Both IR1 and IR2 have counterparts in the 2MASS catalog
and their fluxes from 1.2 to $20.9\,\mu$m
are listed in Table~\ref{tab:ir_sources}.
The spectral energy distributions (SEDs) are rising throughout this
range and, because we do not know where they peak,
our interpretation of their nature is limited. 
A lower limit to their luminosity can be derived by
integrating the portion of each SED that is measured, and gives
$L_{\rm IR1}> 580\,L_\odot, L_{\rm IR2}> 320\,L_\odot$.
A more realistic estimate comes from comparing the fluxes
with the large grid of protostellar models in \cite{2007ApJS..169..328R}.
The 100 lowest $\chi^2$ fits have SEDs that peak between 50 and $100\,\mu$m
and a mean and standard deviation bolometric luminosity,
$L_{\rm IR1}=4200\pm 2200\,L_\odot, L_{\rm IR2}=1600\pm 700\,L_\odot$.
Such high luminosities suggest that IR1 and IR2 are massive protostars
but the uncertainty in how much of the luminosity is due to accretion
prevents us from determining the stellar masses with much precision.

Nevertheless, the diffuse morphology of IR1 at $20\,\mu$m
and its association with both millimeter and centimeter emission
leads us to conclude that it is a warm, dusty HII region embedded
within the cold clump.
The integrated 1.3\,cm VLA flux toward IR1
is 90 mJy which implies a Lyman continuum flux,
$N_{\rm Lyc} = 6\times 10^{46}\,{\rm s}^{-1}$,
that is a factor of 6 less than an O9.5 star
\citep{2005A&A...436.1049M}, and consistent with being an early B star.
The low centimeter luminosity is also comparable to many
more evolved HII regions that have carved out interstellar bubbles
in the Galactic plane \citep{2010ApJ...709..791B}.

IR2 is about half the luminosity of IR1 and a point source in the
near- and mid-infrared.  The lack of centimeter emission indicates
that it is not a strongly ionizing source.
It lies on the edge of the millimeter core and it is therefore hard
to place limits on the amount of cool dust around it but it appears
to have largely dispersed its own protostellar envelope.
We conclude it is likely lower in mass and more evolved than IR1.

The millimeter core itself, MM1, lacks an infrared counterpart at the
limits of detection in the MIRSI image.  Its proximity to the
infrared sources, however, limits how low we can constrain its
luminosity from the Spitzer images (see Figure~\ref{fig:zoom})
but it is certainly lower than the other Spitzer sources in the region.
That is, it has a much lower luminosity than the typical low
mass members of the cluster.  The core mass lies in the range
$12.5-17\,M_\odot$ which is much higher than typical low mass star forming
cores.  Its size and mass imply a surface density $\sim 1$\,g\,cm$^{-2}$
that is on the boundary for high mass star formation
\citep{2008Natur.451.1082K}, but unless the efficiency is very high,
there is simply not enough mass to form an ionizing star and it may
be a progenitor to something more similar to IR2 than IR1.

\subsection{Starless cores in the southern clump}
Although \cite{2006A&A...453.1003T} labeled it as a distinct object,
G111.281--0.670SMM, the southern clump appears more as an elongation
connected to the emission peak in the SCUBA maps.
The SMA filters out most of the extended emission and reveals the compact
cores within this elongated structure (see the lower half of the rightmost
panel in Figure~\ref{fig:zoom}).

We identify three distinct cores, labeled MM3--5 in Figure~\ref{fig:zoom},
with positions and masses given in Table~\ref{tab:mm_cores}.
Their masses are not well determined because of the uncertain contribution
from the shared envelope emission but are in the range typical of
low mass star forming cores ($M_\ast\lesssim 1\,M_\odot$) seen in
Gould Belt clouds \citep{2010ApJ...710.1247S}.
They are all unresolved at the $1\farcs 8\times 1\farcs 5$ resolution
of these data, implying sizes $\lesssim 4000$\,AU,
and therefore rough limits on
column densities $\gtrsim 10^{22}$\,cm$^{-2}$,
and volume densities $\gtrsim 10^6$\,cm$^{-3}$.
These values are high and indicative of objects that are, or shortly will,
form stars.  None of the cores show detectable infrared emission,
however, either in the 2MASS or Spitzer archival images.
Comparing these upper limits to the \cite{2007ApJS..169..328R}.
grid of protostellar models, shows that the total luminosity of
any embedded sources is less than about $80\,L_\odot$,
and these are either very young, deeply embedded, relatively low mass
($M_*\lesssim 2.5\,M_\odot$) protostars or starless cores.
It is tempting to speculate that this region will form
a small chain of stars, somewhat akin to the angled chain of
infrared protostars offset to the west in Figure~\ref{fig:zoom}.

\section{Discussion}
\label{sec:discussion}
It has been well established that high mass stars tend to form in clusters
at the peaks of dense molecular clumps \citep{2007ARA&A..45..481Z}.
Our observations here show two additional luminous objects within a
few arcseconds ($<10^4$\,AU) of the ionizing source of Sh2-157B.
One is a luminous, but non-ionizing, protostar and the other is a
dense millimeter core with mass $\sim 15\,M_\odot$ that is either
starless or contains a deeply embedded object.
In other words, a cluster is being formed but the individual stars
are not born together in perfect synchronicity.

There are also many other lower mass stars visible in the optical and
infrared images of the region, as well as apparently low mass, starless 
cores in the SMA map.  The latter, in particular, are found away from
the main clump peak at lower average surface densities and with a
spatial distribution that follows the elongation of the clump.

The results here resemble our findings from similar
MIRSI and SMA imaging of the ultra-compact HII region AFGL961
in the Rosette molecular cloud \citep{2009ApJ...699.1300W}
where we also found a compact group of protostars in diverse
evolutionary states and a filamentary chain connecting to a starless core.
Several other studies using different tracers and looking at
different objects find comparable results.  For example,
\cite{2007A&A...465..219P} found a cluster of YSOs
in the ultra-compact HII region IRAS 20293+3952
with a similar range of infrared-millimeter colors as found here.
Although we do not discuss spectral line observations here, these
provide many additional ways to delineate evolutionary stages,
through signatures of masers, hot and cold core chemistry,
core collapse, and protostellar outflows
\citep{2007A&A...468.1045B, 2009A&A...499..233F, 2009ApJ...707....1B,
2010A&A...510A...5P, 2011ApJ...739L..16B, 2011A&A...527A..32W}.

To translate evolutionary states into statistical timescales
requires a large systematic survey in much the same way as has been
carried out on low mass, more isolated star forming environments
\citep{2009ApJS..181..321E}.  This would provide a way to
assess how star formation depends on the local environment
and test models of cluster formation.  For example,
if Class 0 cores in and around ultra-compact HII regions
are found at a similar frequency to those in more isolated regions,
then tidal stripping of protostellar envelopes may not
be as important as dynamical models of cluster formation
suggest \citep{2003MNRAS.339..577B}
and would be a point in favor of the picture of quasi-equilibrium
cluster formation \citep{2006ApJ...641L.121T}.

The tight clustering of sources in a range of evolutionary states
show the necessity of arcsecond (or better) resolution images at
multiple wavelengths to gain a more complete picture of cluster birth.
Ground based infrared imaging to follow up on Spitzer obervations plus
sensitive millimeter and centimeter wavelength interferometry provide
a powerful combination and promise a bright future as the ALMA
and Jansky VLA begin operations.

\section{Summary}
\label{sec:summary}
We have carried out a high resolution, multi-wavelength study of the
Sh2-157B region and find evidence for a wide range of protostellar
evolutionary states in close proximity to each other.  The central
ionizing source of the HII region lies at the peak of a massive,
dusty clump.  Our ground-based mid-infrared imaging and millimeter
interferometry show three distinct sources here, all with different
infrared-millimeter colors that suggests the individual components
of a cluster do not form simultaneously but over an extended,
and resolvable, period of time.

Away from the center, we find a chain of three low mass starless cores.
These resemble, at an earlier evolutionary stage, a neighboring chain
of embedded low luminosity infrared sources that also point back towards
the HII region.
On these larger scales, then, we also see that different parts of the
cluster form at different times and perhaps in coherent structures
that derive from the collapsing clump.

These observations are only baby steps toward understanding
star formation in clusters.  They will soon be greatly superseded by
sensitive ALMA observations and, by exploring in more detail the issues
raised here, we can begin to place the subject of clustered, high mass
star formation on a similar level as our current knowledge
of isolated, low mass star formation in nearby clouds.

\acknowledgements
This work is supported by funding from the NSF through grant AST-1108907.
We thank the referee for a careful review, Thomas Robitaille for 
discussions on the interpretation of his protostellar SED fitting software,
and Chris Beaumont for advice on the dendogram structure analysis.
This paper makes use of data products from the Digitized Sky Survey,
the Two Micron All Sky Survey,
the Spitzer Heritage Archive, the VLA archive, and SIMBAD and we acknowledge
the respective project organizations and funding agencies.


\begin{thebibliography}{40}
\expandafter\ifx\csname natexlab\endcsname\relax\def\natexlab#1{#1}\fi

\bibitem[{{Bate} {et~al.}(2003){Bate}, {Bonnell}, \&
  {Bromm}}]{2003MNRAS.339..577B}
{Bate}, M.~R., {Bonnell}, I.~A., \& {Bromm}, V. 2003, \mnras, 339, 577

\bibitem[{{Beaumont} \& {Williams}(2010)}]{2010ApJ...709..791B}
{Beaumont}, C.~N., \& {Williams}, J.~P. 2010, \apj, 709, 791

\bibitem[{{Beuther} {et~al.}(2007){Beuther}, {Zhang}, {Bergin}, {Sridharan},
  {Hunter}, \& {Leurini}}]{2007A&A...468.1045B}
{Beuther}, H., {Zhang}, Q., {Bergin}, E.~A., {Sridharan}, T.~K., {Hunter},
  T.~R., \& {Leurini}, S. 2007, \aap, 468, 1045

\bibitem[{{Bonnell} \& {Bate}(2006)}]{2006MNRAS.370..488B}
{Bonnell}, I.~A., \& {Bate}, M.~R. 2006, \mnras, 370, 488

\bibitem[{{Bonnell} {et~al.}(1998){Bonnell}, {Bate}, \&
  {Zinnecker}}]{1998MNRAS.298...93B}
{Bonnell}, I.~A., {Bate}, M.~R., \& {Zinnecker}, H. 1998, \mnras, 298, 93

\bibitem[{{Brogan} {et~al.}(2011){Brogan}, {Hunter}, {Cyganowski}, {Friesen},
  {Chandler}, \& {Indebetouw}}]{2011ApJ...739L..16B}
{Brogan}, C.~L., {Hunter}, T.~R., {Cyganowski}, C.~J., {Friesen}, R.~K.,
  {Chandler}, C.~J., \& {Indebetouw}, R. 2011, \apjl, 739, L16

\bibitem[{{Brogan} {et~al.}(2009){Brogan}, {Hunter}, {Cyganowski},
  {Indebetouw}, {Beuther}, {Menten}, \& {Thorwirth}}]{2009ApJ...707....1B}
{Brogan}, C.~L., {Hunter}, T.~R., {Cyganowski}, C.~J., {Indebetouw}, R.,
  {Beuther}, H., {Menten}, K.~M., \& {Thorwirth}, S. 2009, \apj, 707, 1

\bibitem[{{Chopinet} \& {Lortet-Zuckermann}(1972)}]{1972A&A....18..373C}
{Chopinet}, M., \& {Lortet-Zuckermann}, M.~C. 1972, \aap, 18, 373

\bibitem[{{Dauphas} \& {Chaussidon}(2011)}]{2011AREPS..39..351D}
{Dauphas}, N., \& {Chaussidon}, M. 2011, Annual Review of Earth and Planetary
  Sciences, 39, 351

\bibitem[{{Evans}(1999)}]{1999ARA&A..37..311E}
{Evans}, II, N.~J. 1999, \araa, 37, 311

\bibitem[{{Evans} {et~al.}(2009){Evans}, {Dunham}, {J{\o}rgensen}, {Enoch},
  {Mer{\'{\i}}n}, {van Dishoeck}, {Alcal{\'a}}, {Myers}, {Stapelfeldt},
  {Huard}, {Allen}, {Harvey}, {van Kempen}, {Blake}, {Koerner}, {Mundy},
  {Padgett}, \& {Sargent}}]{2009ApJS..181..321E}
{Evans}, II, N.~J., {et~al.} 2009, \apjs, 181, 321

\bibitem[{{Fich} \& {Blitz}(1984)}]{1984ApJ...279..125F}
{Fich}, M., \& {Blitz}, L. 1984, \apj, 279, 125

\bibitem[{{Fontani} {et~al.}(2009){Fontani}, {Zhang}, {Caselli}, \&
  {Bourke}}]{2009A&A...499..233F}
{Fontani}, F., {Zhang}, Q., {Caselli}, P., \& {Bourke}, T.~L. 2009, \aap, 499,
  233

\bibitem[{{Hildebrand}(1983)}]{1983QJRAS..24..267H}
{Hildebrand}, R.~H. 1983, \qjras, 24, 267

\bibitem[{{Israel} {et~al.}(1973){Israel}, {Habing}, \& {de
  Jong}}]{1973A&A....27..143I}
{Israel}, F.~P., {Habing}, H.~J., \& {de Jong}, T. 1973, \aap, 27, 143

\bibitem[{{Kassis} {et~al.}(2008){Kassis}, {Adams}, {Hora}, {Deutsch}, \&
  {Tollestrup}}]{2008PASP..120.1271K}
{Kassis}, M., {Adams}, J.~D., {Hora}, J.~L., {Deutsch}, L.~K., \& {Tollestrup},
  E.~V. 2008, \pasp, 120, 1271

\bibitem[{{Krumholz} {et~al.}(2009){Krumholz}, {Klein}, {McKee}, {Offner}, \&
  {Cunningham}}]{2009Sci...323..754K}
{Krumholz}, M.~R., {Klein}, R.~I., {McKee}, C.~F., {Offner}, S.~S.~R., \&
  {Cunningham}, A.~J. 2009, Science, 323, 754

\bibitem[{{Krumholz} \& {McKee}(2008)}]{2008Natur.451.1082K}
{Krumholz}, M.~R., \& {McKee}, C.~F. 2008, \nat, 451, 1082

\bibitem[{{Kurtz} {et~al.}(1994){Kurtz}, {Churchwell}, \&
  {Wood}}]{1994ApJS...91..659K}
{Kurtz}, S., {Churchwell}, E., \& {Wood}, D.~O.~S. 1994, \apjs, 91, 659

\bibitem[{{Kurtz} {et~al.}(1999){Kurtz}, {Watson}, {Hofner}, \&
  {Otte}}]{1999ApJ...514..232K}
{Kurtz}, S.~E., {Watson}, A.~M., {Hofner}, P., \& {Otte}, B. 1999, \apj, 514,
  232

\bibitem[{{Lada} \& {Lada}(2003)}]{2003ARA&A..41...57L}
{Lada}, C.~J., \& {Lada}, E.~A. 2003, \araa, 41, 57

\bibitem[{{Martins} {et~al.}(2005){Martins}, {Schaerer}, \&
  {Hillier}}]{2005A&A...436.1049M}
{Martins}, F., {Schaerer}, D., \& {Hillier}, D.~J. 2005, \aap, 436, 1049

\bibitem[{{McKee} \& {Ostriker}(2007)}]{2007ARA&A..45..565M}
{McKee}, C.~F., \& {Ostriker}, E.~C. 2007, \araa, 45, 565

\bibitem[{{McKee} \& {Tan}(2003)}]{2003ApJ...585..850M}
{McKee}, C.~F., \& {Tan}, J.~C. 2003, \apj, 585, 850

\bibitem[{{McKee} \& {Williams}(1997)}]{1997ApJ...476..144M}
{McKee}, C.~F., \& {Williams}, J.~P. 1997, \apj, 476, 144

\bibitem[{{Palau} {et~al.}(2007){Palau}, {Estalella}, {Girart}, {Ho}, {Zhang},
  \& {Beuther}}]{2007A&A...465..219P}
{Palau}, A., {Estalella}, R., {Girart}, J.~M., {Ho}, P.~T.~P., {Zhang}, Q., \&
  {Beuther}, H. 2007, \aap, 465, 219

\bibitem[{{Palau} {et~al.}(2010){Palau}, {S{\'a}nchez-Monge}, {Busquet},
  {Estalella}, {Zhang}, {Ho}, {Beltr{\'a}n}, \&
  {Beuther}}]{2010A&A...510A...5P}
{Palau}, A., {S{\'a}nchez-Monge}, {\'A}., {Busquet}, G., {Estalella}, R.,
  {Zhang}, Q., {Ho}, P.~T.~P., {Beltr{\'a}n}, M.~T., \& {Beuther}, H. 2010,
  \aap, 510, A5

\bibitem[{{Robitaille} {et~al.}(2007){Robitaille}, {Whitney}, {Indebetouw}, \&
  {Wood}}]{2007ApJS..169..328R}
{Robitaille}, T.~P., {Whitney}, B.~A., {Indebetouw}, R., \& {Wood}, K. 2007,
  \apjs, 169, 328

\bibitem[{{Rosolowsky} {et~al.}(2008){Rosolowsky}, {Pineda}, {Kauffmann}, \&
  {Goodman}}]{2008ApJ...679.1338R}
{Rosolowsky}, E.~W., {Pineda}, J.~E., {Kauffmann}, J., \& {Goodman}, A.~A.
  2008, \apj, 679, 1338

\bibitem[{{Sadavoy} {et~al.}(2010){Sadavoy}, {Di Francesco}, {Bontemps},
  {Megeath}, {Rebull}, {Allgaier}, {Carey}, {Gutermuth}, {Hora}, {Huard},
  {McCabe}, {Muzerolle}, {Noriega-Crespo}, {Padgett}, \&
  {Terebey}}]{2010ApJ...710.1247S}
{Sadavoy}, S.~I., {et~al.} 2010, \apj, 710, 1247

\bibitem[{{Sharpless}(1959)}]{1959ApJS....4..257S}
{Sharpless}, S. 1959, \apjs, 4, 257

\bibitem[{{Tan} {et~al.}(2006){Tan}, {Krumholz}, \&
  {McKee}}]{2006ApJ...641L.121T}
{Tan}, J.~C., {Krumholz}, M.~R., \& {McKee}, C.~F. 2006, \apjl, 641, L121

\bibitem[{{Thompson} {et~al.}(2006){Thompson}, {Hatchell}, {Walsh},
  {MacDonald}, \& {Millar}}]{2006A&A...453.1003T}
{Thompson}, M.~A., {Hatchell}, J., {Walsh}, A.~J., {MacDonald}, G.~H., \&
  {Millar}, T.~J. 2006, \aap, 453, 1003

\bibitem[{{Wang} {et~al.}(2011){Wang}, {Beuther}, {Bik}, {Vasyunina}, {Jiang},
  {Puga}, {Linz}, {Rod{\'o}n}, {Henning}, \& {Tamura}}]{2011A&A...527A..32W}
{Wang}, Y., {et~al.} 2011, \aap, 527, A32

\bibitem[{{Williams}(2010)}]{2010ConPh..51..381W}
{Williams}, J. P. 2010, Contemporary Physics, 51, 381

\bibitem[{{Williams} {et~al.}(2000){Williams}, {Blitz}, \&
  {McKee}}]{2000prpl.conf...97W}
{Williams}, J.~P., {Blitz}, L., \& {McKee}, C.~F. 2000, Protostars and Planets
  IV, 97

\bibitem[{{Williams} {et~al.}(1994){Williams}, {de Geus}, \&
  {Blitz}}]{1994ApJ...428..693W}
{Williams}, J.~P., {de Geus}, E.~J., \& {Blitz}, L. 1994, \apj, 428, 693

\bibitem[{{Williams} {et~al.}(2009){Williams}, {Mann}, {Beaumont}, {Swift},
  {Adams}, {Hora}, {Kassis}, {Lada}, \&
  {Rom{\'a}n-Z{\'u}{\~n}iga}}]{2009ApJ...699.1300W}
{Williams}, J.~P., {et~al.} 2009, \apj, 699, 1300

\bibitem[{{Wolfire} \& {Cassinelli}(1987)}]{1987ApJ...319..850W}
{Wolfire}, M.~G., \& {Cassinelli}, J.~P. 1987, \apj, 319, 850

\bibitem[{{Zinnecker} \& {Yorke}(2007)}]{2007ARA&A..45..481Z}
{Zinnecker}, H., \& {Yorke}, H.~W. 2007, \araa, 45, 481

\end{thebibliography}
\end{document}